\documentclass[prl,aps,showpacs,twocolumn,superscriptaddress,nobibnotes,epsf]{revtex4}

\usepackage{graphicx}
\usepackage{dcolumn}
\usepackage{bm}
\usepackage{SIunits}

\begin{document}

\title{Enhanced Superconductivity by Rare-earth Metal-doping in Phenanthrene}

\author{X. F. Wang}
\author{X. G. Luo}
\author{J. J. Ying}
\author{Z. J. Xiang}
\affiliation{Hefei National Laboratory for Physical Science at
Microscale and Department of Physics, University of Science and
Technology of China, Hefei, Anhui 230026, People's Republic of
China}
\author{S. L. Zhang}
\author{R. R. Zhang}
\affiliation{High Magnetic Field Laboratory, Chinese Academy of
Sciences, Hefei 230031, People¡¯s Republic of China}
\author{Y. H. Zhang}
\affiliation{Hefei National Laboratory for Physical Science at
Microscale and Department of Physics, University of Science and
Technology of China, Hefei, Anhui 230026, People's Republic of
China}
 \affiliation{High Magnetic Field Laboratory, Chinese Academy of Sciences, Hefei 230031, People¡¯s Republic of China}
\author{Y. J. Yan}
\author{A. F. Wang}
\author{P. Cheng}
\author{G. J. Ye}
\author{X. H. Chen}
\altaffiliation{Corresponding author} \email{chenxh@ustc.edu.cn}
\affiliation{Hefei National Laboratory for Physical Science at
Microscale and Department of Physics, University of Science and
Technology of China, Hefei, Anhui 230026, People's Republic of
China}

\begin{abstract}
We successfully synthesized La- and Sm-doped phenanthrene powder
samples and discovered superconductivity at $T_{\rm c}$ around 6 K
in them. The $T_{\rm c}$s are 6.1 K for LaPhenanthrene and 6.0 K for
SmPhenanthrene, which are enhanced by about 1 K and 0.5 K compared
to those in $A_3$Phenanthrene ($A$=K and Rb) and in
$Ae_{1.5}$Phenanthrene ($Ae$ = Sr and Ba) superconductors
respectively. The superconductive shielding fractions for
LaPhenanthrene and SmPhenanthrene are 46.1$\%$ and 49.8$\%$ at 2 K,
respectively. The little effect of the doping of the magnetic ion
Sm$^{3+}$ on $T_c$ and the positive pressure dependence coefficient
on $T_{\rm c}$ strongly suggests unconventional superconductivity in
the doped phenanthrene superconductors. The charge transfer to
organic molecules from dopants of La and Sm induces a redshift of 7
cm$^{-1}$ per electron for the mode at 1441 cm$^{-1}$ in the Raman
spectra, which is almost the same as those observed in
$A_3$Phenanthrene ($A$=K and Rb) and $Ae_{1.5}$Phenanthrene ($Ae$ =
Sr and Ba) superconductors.

\end{abstract}
\pacs{74.70.Kn,74.62.Fj,74.25.-q}
\vskip 300 pt

\maketitle

The first observation of superconductivity in a carbon-based
compound traces back to 1965, when superconductivity was observed in
the first stage alkali-metal intercalated graphite C$_8$K.\cite{ck}
Up to now, the superconductors based on carbon mainly consist of
three types of materials: graphite intercalation compounds (GIC),
doped fullerenes, and organic compounds. For all of these
carbon-based superconductors, there are commonly five-member rings
or six member rings with conjugated $\pi$-orbital interactions among
these rings. The $\pi$-electron can delocalize throughout the
crystal, giving rise to metallic conductivity due to a $\pi$-orbital
overlap between the adjacent molecules.  At present, the
highest-$T_{\rm c}$ superconductor for these carbon-based materials
is an alkali metal-doped fullerene, namely Cs$_3$C$_{60}$ under
pressure ($\sim$ 12 kbar) with $T_c\sim$38 K.\cite{CsC60,CsC60-2}
Among the organic compounds, the previous record of $T_{\rm c}$ was
held by the tetrathiafulvalene derivative
(BEDT-TTF)$_2$CuN(CN)$_2$Cl with $T_{\rm c}$=12.8 K under 0.3 kbar
pressure\cite{BEDT-HP}. While, very recently, this record of $T_{\rm
c}$ among superconducting organic materials was renewed by the
patassium-doped picene (highest $T_c\sim$18K)\cite{SC-picene} and
subsequently the potassium-doped 1,2:8,9-dibenzopentacene ($T_{\rm
c}\sim$31 K)\cite{Dp}, whose pristine organic molecules compose of
five and seven fused benzene rings, respectively. Alkali and alkali
earth metal-doped phenanthrenes, whose pristine organic molecule
phenanthrene consists of three fused benzene rings, are also found
be superconducting with $T_{\rm c}\sim$ 5 K\cite{xfwang, xfwang2}.
These discoveries of superconductivity in the materials with the
fused benzene rings could open a novel broad class of hydrocarbon
organic materials for superconductors, and suggest the potential
high-$T_{\rm c}$ superconductivity in the materials with long fused
benzene rings. However, the mechanism for these organic hydrocarbons
with long benzene rings still remain open. Some traces have been
observed to hint an unconventional superconductivity in these
materials, such as the positive pressure dependence of $T_{\rm c}$
and the existence of local spin moments in the superconducting
compounds\cite{xfwang, xfwang2}. More detailed work should be deeply
conducted on this class of superconductors. The superconductivity in
doped phenanthrene offers an good candidate for investigating the
physics in such organic superconductors, due to the relatively
simple molecular structure of phenanthrene. In this work, we doped
non-magnetic and magnetic rare-earth metal elements into the
phenanthrene to study the effect of magnetic ions on the
superconductivity. Superconductivity in this class of hydrocarbon
organic materials was realized by the doping-induced charge
(electron) transfer from the doped metal atoms to the molecules,
which results in changes of the electronic structure and the
physical properties. Our previous works demonstrated that 3
electrons are required to transfer into one phenanthrene molecule to
get superconductivity. In the present letter, we doped rear-earth
metals lanthanum and samarium, which are non-magnetic and magnetic
respectively, into phenanthrene to realize superconductivity, with
nominal composition of La$_1$Phenanthrene and Sm$_1$Phenanthrene.
Superconductivity was observed in these two compounds, with $T_{\rm
c}$ equal to 6.1 K and 6.0 K for LaPhenanthrene and SmPhenanthrene,
respectively. The superconductive shielding fraction is 46.1$\%$ in
LaPhenanthrene and 49.8$\%$ in SmPhenanthrene at 2.5 K. Raman
spectra show a redshift of $\sim$ 7 cm$^{-1}$ per electron due to
the charge transfer, which is almost the same as those in
$A_3$Phenanthrene ($A$=K and Rb) and $Ae_{1.5}$Phenanthrene ($Ae$=Ba
and Sr) as well as that in $A_3$C$_{60}$ ($A$=K and Rb). The
pressure dependence of superconductivity in LaPhenanthrene shows a
positive coefficient d\emph{(T}$_{\rm c}$\emph{/T}$_{\rm
c}$(0)\emph{)}/d\emph{P} with the superconductive shielding fraction
almost unchanged.

\begin{figure}[h]
\centering
\includegraphics[width=0.48\textwidth]{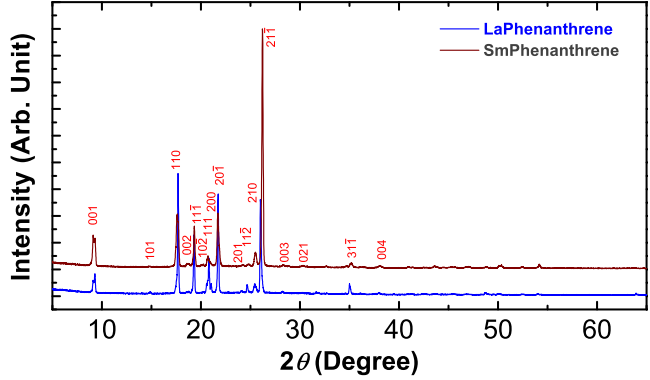}
\caption{X-Ray diffraction patterns for the superconducting
LaPhenanthrene and SmPhenanthrene. The LaPhenanthrene and SmPhenanthrene
crystallizes into the same structure as the pristine phenanthrene with space group \emph{$P2_1$}.} \label{fig1}
\end{figure}

The phenanthrene(98$\%$) was purified by sublimation method
\cite{xfwang}. Lanthanum(99.99$\%$) and samarium(99.99$\%$) were
ground into powder with file and mixed carefully with the purified
phenanthrene in molar ratio of 1:1 respectively. The synthesis of
LaPhenanthrene and SmPhenanthrene is quite similar to the processes
for Ba- and Sr-doped phenanthrene \cite{xfwang2}. The mixture of
rare-earth metal and phenanthrene was sintered at 240$\celsius$ with
the multiple-middle-treating process for totally 8 days. Finally,
the products with uniform dark black color were obtained. The X-ray
diffraction and Raman measurement were carried out by sealing the
samples in capillaries made of special glass No. 10 and purchased
from Hilgenberg GmbH. X-ray diffraction pattern was obtained in the
2-theta range of 5$\degree$¨C65$\degree$ with a scanning rate of
0.5$\degree$ per minute. Raman-scattering experiments were conducted
by using the 780-nm laser line in the DXR Raman Microscope ( Thermo
Scientific ). The scattering light was captured by using a single
exposure of the CCD with a spectral resolution of 1 cm$^{-1}$.
Low-temperature Raman spectra were performed on a Raman Microscope
(Horiba JY T64000) equipped with Janis ST-500 Microscopy cryostat.

Figure 1 shows the X-ray diffraction (XRD) patterns of the
La- and Sm-doped phenanthrene. The pristine phenanthrene crystallizes in the space group of
\emph{P$_2$$_{1}$}.\cite{trotter,xfwang} The lattice parameters for the pristine
phenanthrene are $a$ = 8.453${\rm \AA}$, $b$ = 6.175${\rm \AA}$, $c$ = 9.477${\rm \AA}$ and
$\beta$=98.28$^{\degree}$.\cite{trotter,xfwang} All the reflections in the XRD patterns shown in Fig. 1 can be well indexed
with the space group of \emph{P}$2$$_{\emph{1}}$ as that for the pristine phenanthrene, just the same as previously reported
alkali and alkali-earth metal-doped phenanthrene. No impurity phase was found in the XRD patterns. From the XRD patterns shown
in Fig. 1, lattice parameters are obtained as  $a$ = 8.481
${\rm \AA}$, $b$ = 6.187${\rm \AA}$, $c$= 9.512${\rm \AA}$, $\beta$ = 97.95$^{\degree}$ for
LaPhenanthrene and $a$ = 8.475${\rm \AA}$, $b$ = 6.180${\rm \AA}$, $c$ =
9.505${\rm \AA}$, $\beta$ = 98.10$^\degree$ for SmPhenanthrene. The lattice parameters change
slightly relative to the pristine phenanthrene. The unit cell
volume expands from 489.5 ${\rm \AA}^3$ for the pristine phenanthrene to
494.3 ${\rm \AA}^3$ for LaPhenanthrene and 492.9 ${\rm \AA}^3$ for
SmPhenanthrene. The expansion of the unit cell is similar to the previous alkali-earth metal-doped case.
The smaller unit cell of SmPhenanthrene relative to LaPhenanthrene is consistent with the smaller ion radius of Sm$^{3+}$ than that of La$^{3+}$.

\begin{figure}[h]
\centering
\includegraphics[width=0.48\textwidth]{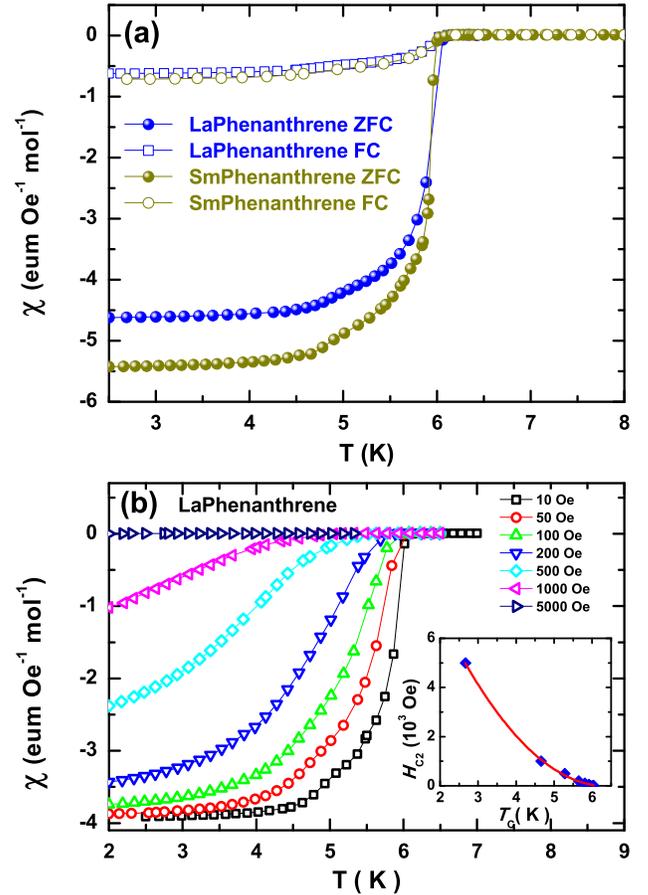}
\caption{Temperature dependence of magnetic susceptibility ($\chi$) for
LaPhenanthrene and SmPhenanthrene. (a). $\chi$
plotted against $T$ for LaPhenanthrene and SmPhenanthrene
in the zero-field-cooling (ZFC) and field-cooling (FC) measurements
under the magnetic field of 10 Oe. (b). Magnetic susceptibility as a function of temperature for
LaPhenanthrene in the ZFC measurements under different
magnetic fields. The $H$ versus $T_{\rm c}$ is plotted in the inset of (b).}
\label{fig2}
\end{figure}

\begin{figure}[h]
\centering
\includegraphics[width=0.48\textwidth]{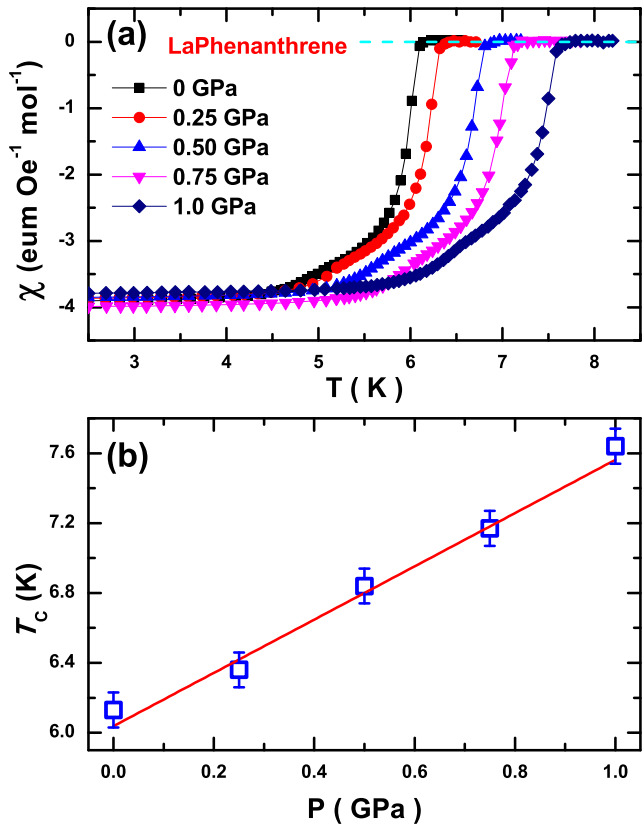}
\caption{Pressure dependence of superconducting transition temperature
\emph{T}$_{\rm c}$ for LaPhenanthrene. (a). Magnetic susceptibility plotted against \emph{T} in
ZFC measurements for LaPhenanthrene under the pressures of $P$
= 0, 0.25, 0.6, 0.8 and 1.0 GPa. (b). \emph{T}$_c$ plotted as a function of pressure for the superconducting LaPhenanthrene.} \label{fig3}
\end{figure}

Superconductivity of the LaPhenanthrene and SmPhenanthrene powder samples was characterized by magnetic susceptibility measurements
at low magnetic field in zero-field-cooling (ZFC) and field-cooling (FC) processes. Figure 2(a) displays the temperature dependence of magnetic
susceptibility $\chi$($T$) measured in the ZFC and FC processes and under the magnetic field of 10 Oe for the powder samples of
 LaPhenanthrene and SmPhenanthrene. $\chi$($T$)s exhibit clear and sharp drops in ZFC and FC measurements for both LaPhenanthrene
 and SmPhenanthrene at the temperatures of 6.1 K and 6.0 K, respectively. The temperature for the beginning of the sharp drop is
 defined as the superconducting transition temperature ($T_{\rm c}$). Diamagnetic signals from ZFC and FC measurements can be
 assigned to superconductive shielding and Meissner effect. As shown in Fig.2(a), the shielding fraction and the Meissner fraction
 are 46.1\% and 6.3\% for the powder sample of LaPhenanthrene, 49.8\% and 6.5\% for the sample of SmPhenanthrene, respectively.
 The shielding fraction is much larger than that of alkali-metal doped picene \cite{SC-picene} and phenanthrene \cite{xfwang}.
 The $T_{\rm c}$s for these two rare-earth metal-doped phenanthrene
 are slightly higher than those of alkali metal-doped (less than 5 K) and alkali-earth metal-doped phenanthrene ($\sim$ 5.5 K), respectively.

Temperature dependence of the magnetic susceptibility of the
LaPhenanthrene superconductor was measured under different magnetic
fields in ZFC processes, as displayed in Fig. 2(b) . The diamagnetic
signal becomes smaller and the superconducting transition gets
broader with increasing the magnetic field. Superconducting
transition can still be observed at 2.7 K under the magnetic field
of 5000 Oe. When the field is higher than 8000 Oe, no
superconducting transition is observed down to 2 K in magnetic
susceptibility. The relationship between $T_{\rm c}$ and the applied
magnetic field $H$ is plotted in the inset of Fig.2(b). It is
difficult to precisely determine the zero-temperature-limiting upper
critical field \emph{H}$_{\rm C2}$(0) from \emph{H-T$_{\rm c}$}
plot, but it must higher than 5000 Oe.

Figure 3(a) shows the magnetic susceptibility of LaPhenanthrene in
ZFC measurements as a function of temperature at different pressures
up to 1 GPa.   The superconducting transition temperature increases
from 6.1 K to 7.6 K as the pressure increases form ambient pressure
to 1 GPa while remain the magnitude of the diamagnetic signal almost
unchanged. Figure 3(b) exhibits the pressure dependence of $T_{\rm
c}$. From Fig. 3(b), the temperature dependence coefficient of
$T_{\rm c}$, d\emph{(T}$_{\rm c}(P)$\emph{/T}$_{\rm
c(0)})$/d\emph{P}, can be estimated as $\sim$ 0.25GPa$^{-1}$. This
value is nearly the same as those in Sr$_{1.5}$Phenanthrene
($\sim$0.21 GPa$^{-1}$) and in Ba$_{1.5}$Phenanthrene ($\sim$0.23
GPa$^{-1}$),\cite{xfwang2} and in K$_3$Phenanthrene ($\sim$0.26
GPa$^{-1}$).\cite{xfwang} The positive pressure effect of $T_{\rm
c}$ is the common feature in the doped phenanthrene superconductors.
In BCS theory, the suppression effect of \emph{T}$_{\rm c}$ is
expected as an external pressure is applied because the external
pressure compresses the crystal lattice and consequently reduces the
density of states at the Fermi level. Therefore, the unusual
positive pressure effect of $T_{\rm c}$ in the doped phenanthrene
superconductors could be a strong evidence for unconventional
superconductivity in the doped phenanthrene superconductors.

\begin{figure}[h]
\centering
\includegraphics[width=0.48\textwidth]{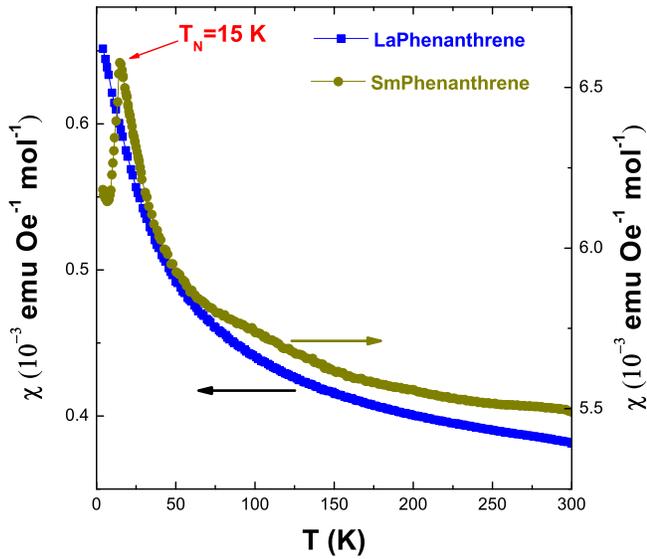}
\caption{High-field magnetic susceptibility for the
superconducting LaPhenanthrene and SmPhenanthrene. }
\label{fig4}
\end{figure}

\begin{figure}[h]
\centering
\includegraphics[width=0.48\textwidth]{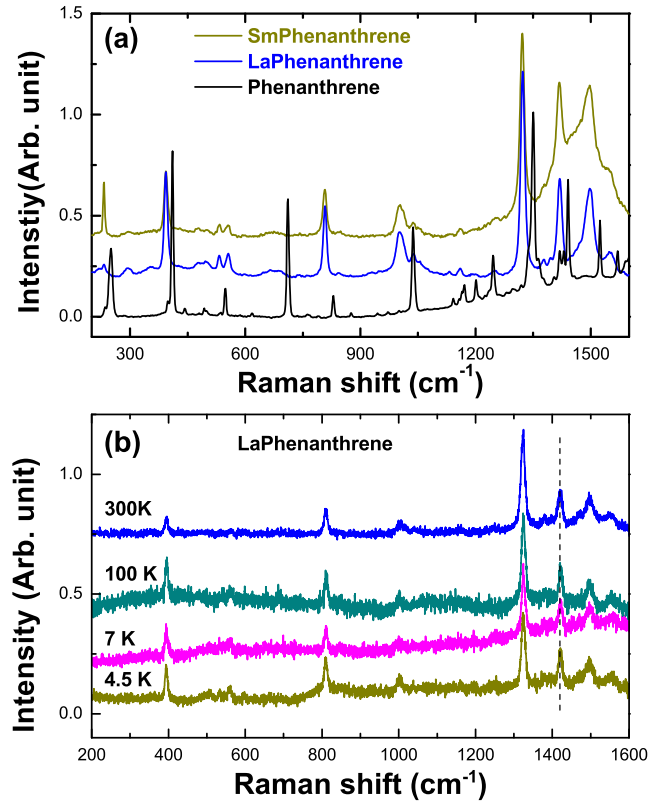}
\caption{(a) Raman spectra taken at room temperature for the pure phenanthrene, the
superconducting LaPhenanthrene and SmPhenanthrene. (b) Raman spectra for LaPhenanthrene taken at different temperature.}
\label{fig5}
\end{figure}

The temperature dependence of the magnetic susceptibility measured
at 5 T for LaPhenanthrene and SmPhenanthrene powder samples is
displayed in Fig. 4. Curie-Weiss-like behavior is observed at high
temperature for both samples, indicating the existence of local spin
moments. Especially, for the SmPhenanthrene, there is an
antiferromagnetic transition happening at $T_{\rm N}\sim$15 K, which
is ascribed to the antiferromagnetic ordering of Sm$^{3+}$ ions.
This $T_{\rm N}$ is close to that in the elemental Sm
(14.8K).\cite{Sm0} But no reflection for elemental Sm can be
recognized in the XRD pattern shown in Fig. 1. Actually, $T_{\rm N}$
for Sm$^{3+}$ ions can vary from several kelvins to higher than 20
K, which strongly depends on concrete compounds.\cite{Sm1,Sm2,Sm3}
While it was reported that no magnetic order is formed for Sm$^{3+}$
ions In Sm$_{2.75}$C$_{60}$.\cite{Sm4} One can find that the
existence of the magnetic Sm$^{3+}$ ions with antiferromagnetic
order has negligible suppression effect on the superconductivity. It
strongly suggests the unconventional nature of the superconductivity
in the system.

As one knows, the effect of charge transferring into the fused
hydrocarbon rings can be studied by Raman spectroscopies. Figure
5(a) shows the Raman spectra at room temperature for the pristine
phenanthrene, LaPhenanthrene and SmPhenanthrene. As reported
previously, seven major peaks can be observed in the pristine
phenanthrene: 1524, 1441, 1350, 1037, 830, 411 and 250 cm$^{-1}$.
All of these major peaks of the Raman spectrum can classified to
$A_{1}$ mode, which arises from the C-C stretching
vibrations\cite{Martin,Godec}. Clear downshift can be observed for
the peaks in the Raman spectra of LaPhenanthrene and SmPhenanthrene,
relative to those of the pristine phenanthrene. Such phonon-mode
softening effect is attributed to the charge transfer from dopants
of La and Sm atoms into phenanthrene molecules, which has been
widely observed in the doped fullerenes.\cite{Fullerene} Kato et al.
pointed out theoretically that in monoanion of phenanthrene the
stretching $A_1$ mode of 1434 cm$^{-1}$ strongly couples to the
$a_2$ lowest unoccupied molecular orbital (LUMO).\cite{kato} This
suggests that this peak (1441 cm$^{-1}$ in the pristine
phenanthrene) can be most affected by charge transfer to the
molecules and the downshift wavenumber for one electron transfer is
7 cm$^{-1}$. For the present two rare-earth metal-doped phenanthrene
superconductors, this $A_1$ mode of 1441 cm$^{-1}$ in the pristine
phenanthrene moves to 1420 cm$^{-1}$ in LaPhenanthrene and 1419
cm$^{-1}$ in SmPhenanthrene. Consequently, the electron transfer
from rare-earth metal atoms to the phenanthrene molecules induces
redshifts of 21 and 22 cm$^{-1}$ in Raman spectra for LaPhenanthrene
and SmPhenanthrene respectively, which corresponds to redshift of
about 7 cm$^{-1}$ per electron in both samples. This value of
redshift per electron transfer is consistent with those observed in
the alkali and alkali earth metal-doped phenanthrene superconductors
(6-8 cm$^{-1}$ per electron) and also consistent with the
theoretical prediction by Kato $et~al.$.\cite{kato} Temperature
dependent measurements of Raman scattering were also performed for
LaPhenanthrene at the temperatures ranging from 300 K to 4.5 K. The
selected Raman spectra are displayed in Fig. 5(b). Normally, phonon
anharmonicity gives rise to the following temperature dependence of
the linewidth (full width of half-maximum) $\Gamma$ and mode
frequency $\omega$:\cite{RamanT}
\begin{equation} \Gamma(T) = \gamma +\Gamma_0\left(1+\frac{2}{e^{2\hbar\omega_0/k_{\rm B}T}-1}\right) \end{equation}
\begin{equation} \omega(T) = \omega_0 +C\left(1+\frac{2}{e^{2\hbar\omega_0/k_{\rm B}T}-1}\right) \end{equation}
where $\hbar\omega_0$ is the phonon frequency, $k_{\rm B}$ is the Boltzmann constant, and C is a constant. In Fig. 5(b),
no shift more than 0.5 cm$^{-1}$ can be observed for both linewidth and mode frequency with the variation of the temperature
 from 300 K to 4.5 K, indicating an unusual phonon decay with decreasing temperature. The nearly temperature independent
 $\Gamma(T)$ suggests $\Gamma_0\ll\gamma$ so that lattice anharmonicity is no long a dominant mechanism of the phonon decay
 in the present studied LaPhenanthrene. The mechanism for this unusual process of phonon decay has not been understood yet.
 The interactions of phonon with other excitations (for instance, electrons in conducting materials), might affect the phonon
 lifetime and contribute to the phonon linewidth $\Gamma(T )$ because additional channels for phonon decay could be opened by
 them. More measurements with higher resolution must be done to explain the temperature-independent linewidth and mode frequency.
 It should also be pointed out that, as seen from Fig. 5b, the Raman spectrum exhibits no obvious change as
 temperature cools down across $T_{\rm c}$. However, all of these mode frequencies are much larger than 2$\Delta\sim$ 15 cm$^{-1}$
 (assume a BCS gap energy 2$\Delta$ = 3.53$k_{\rm B}T_{\rm c}$) and so that most of these modes are not expected to be sensitive
 to the superconducting transition.

In summary, we successfully fabricated the new superconductors
LaPhenanthrene and SmPhenanthren, which show $T_{\rm c}$ = 6.1 and
6.0 K, respectively. The positive pressure effect on $T_{\rm c}$ was
observed in LaPhenanthrene. Especially, the antiferromagnetic order
from the magnetic ion Sm$^{3+}$ has negligible effect on $T_{\rm c}$
in SmPhenanthren. These results strongly suggest unconventional
superconductivity in the doped phenanthrene-type superconductors.

\begin{acknowledgments}
This work is supported by National Natural Science Foundation of China (Grant No. 11190021, 51021091), the National Basic Research
Program of China (973 Program, Grant No. 2012CB922002 and No. 2011CB00101), and Chinese Academy of Sciences.

\end{acknowledgments}

\end{document}